\documentclass[journal = ancac3, manuscript = article, layout = twocolumn]{achemso}

\usepackage{graphicx}
\usepackage[]{xspace}

\def\mose2{MoSe$_2$\xspace}
\def\wse2{WSe$_2$\xspace}
\def\rc{$\mu$RC\xspace}
\def\pl{$\mu$PL\xspace}

\def\ya{\textbf{y}-axis\xspace}

\title{Exciton band structure in layered MoSe$_2$: From monolayer to bulk limit}
\author{Ashish Arora}
\email{ashish.arora@lncmi.cnrs.fr}
\affiliation{Laboratoire National des Champs Magn\'{e}tiques Intenses\\ CNRS-UJF-UPS-INSA, 25 rue des Martyrs, 38042 Grenoble, France}
\altaffiliation{Current address: Physikalisches Institut, Westf\"{a}lische Wilhelms-Universit\"{a}t M\"{u}nster, Wilhelm-Klemm-Stra{\ss}e 10, 48149 M\"{u}nster, Germany}
\author{Karol Nogajewski}
\affiliation{Laboratoire National des Champs Magn\'{e}tiques Intenses\\ CNRS-UJF-UPS-INSA, 25 rue des Martyrs, 38042 Grenoble, France}
\author{Maciej R. Molas}
\affiliation{Laboratoire National des Champs Magn\'{e}tiques Intenses\\ CNRS-UJF-UPS-INSA, 25 rue des Martyrs, 38042 Grenoble, France}
\alsoaffiliation{Institute of Experimental Physics,
Faculty of Physics, University of Warsaw, Pasteura 5, 02-093 Warszawa, Poland}
\author{Maciej Koperski}
\affiliation{Laboratoire National des Champs Magn\'{e}tiques Intenses\\ CNRS-UJF-UPS-INSA, 25 rue des Martyrs, 38042 Grenoble, France}
\alsoaffiliation{Institute of Experimental Physics,
Faculty of Physics, University of Warsaw, Pasteura 5, 02-093 Warszawa, Poland}
\author{Marek Potemski}
\email{marek.potemski@lncmi.cnrs.fr}
\affiliation{Laboratoire National des Champs Magn\'{e}tiques Intenses\\ CNRS-UJF-UPS-INSA, 25 rue des Martyrs, 38042 Grenoble, France}


\begin{document}
\begin{abstract}
\textbf{We present the micro-photoluminescence (\pl) and micro-reflectance contrast (\rc) spectroscopy studies on thin films of \mose2 with layer thicknesses ranging from a monolayer (1L) up to 5L. The thickness dependent evolution of the ground and excited state excitonic transitions taking place at various points of the Brillouin zone is determined. Temperature activated energy shifts and linewidth broadenings of the excitonic resonances in 1L, 2L and 3L flakes are accounted for by using standard formalisms previously developed for semiconductors. A peculiar shape of the optical response of the ground state (A) exciton in monolayer \mose2 is tentatively attributed to the appearance of Fano-type resonance. Rather trivial and clearly decaying PL spectra of monolayer \mose2 with temperature confirm that the ground state exciton in this material is optically bright in contrast to a dark exciton ground state in monolayer \wse2.}
\vspace{0.5cm}
\end{abstract}
\section{\label{secIntro}Introduction}
Layered semiconducting transition metal dichalcogenides (TMDCs) such as MoS$_2$, MoSe$_2$, WS$_2$, WSe$_2$ and MoTe$_2$ have become a subject of intensive study recently due to their interesting optical properties rich in spin and valley physics, which largely results from the indirect to direct band gap crossover upon decreasing thickness from bulk to the monolayer (1L).\cite{prl10km,nn12qhw,an13ge,an13szb,prl12dx,oe13pt,nl15igl} Thin layers of these materials have shown new interesting physical phenomena such as valley polarization\cite{nc12tc,nn12kfm} and valley coherence\cite{nn13amj} effects or tightly bound trions.\cite{nm13kfm,nc13jsr} It has also been domenstrated that they can serve as efficient sources of light, either in the form of light emitting diodes\cite{nn14ap} or single photon emitters,\cite{nn15mk} which is potentially promising towards technological advances such as in optical computation and valleytronics. As such, the optical properties of TMDCs are dominated by many excitonic resonances over a rather small spectral region (a few eV). Therefore, a full comprehensive picture of the electronic band structure requires a detailed experimental study of excitons, and their excited states as a function of layer thickness and temperature, which has received very limited attention so far. In this letter, we present a comprehensive analysis of the evolution of many excitonic ground state and excited state features, spanning the \rc spectra of good quality 1L to 5L thick flakes of \mose2. This information finds importance for an improved theoretical understanding of excitons and charged excitons where large discrepancies have been found between the experimental and limited theoretical results (such as exciton binding energies).\cite{prb12ar,prl13dyq,arxiv15mzm} Interestingly, for the A-exciton transition in our monolayer flakes, we find an unusual reflectance spectral lineshape, which is tentatively attributed to the appearance of a Fano-type resonance.\cite{pr61uf,prb94dyo,sst96vb,sse96vb,ssc96vb} This could arise since the A-exciton lies in the energy region spanned by the excited and the continuum states of the trion, where a strong interaction could exist.\cite{pssb01ae,pssb01ras} Temperature dependent evolution of the \rc spectrum of the monolayer flake is consistent with the Fano hypothesis. Recent theoretical works suggest that in addition to strong spin-orbit splitting in the valence band (VB) of TMDCs, there is also a splitting in the conduction band (CB) (up to $\sim$40~meV).\cite{prb13kk,2dm15ak,arxiv15hd} The sign of this splitting is expected to decide weather the energetically favorable exciton is optically active (bright) or inactive (dark) which has a strong influence on the PL properties of the material. Our temperature-dependent \pl studies on the monolayer \mose2 are consistent with the theoretical predictions that the ground state exciton in this material is optically bright, which is found to be opposite to the case of \wse2 monolayers.\cite{ns15aa} We also derive the temperature activated energy shifts and linewidth broadenings of the ground state excitons from 1L to 3L thick flakes and explain them on the basis of standard formalisms used for conventional semiconductors.

\section{\label{secCharacterization}Samples and characterization}
\begin{figure}[!t]
\includegraphics[width=8.5cm]{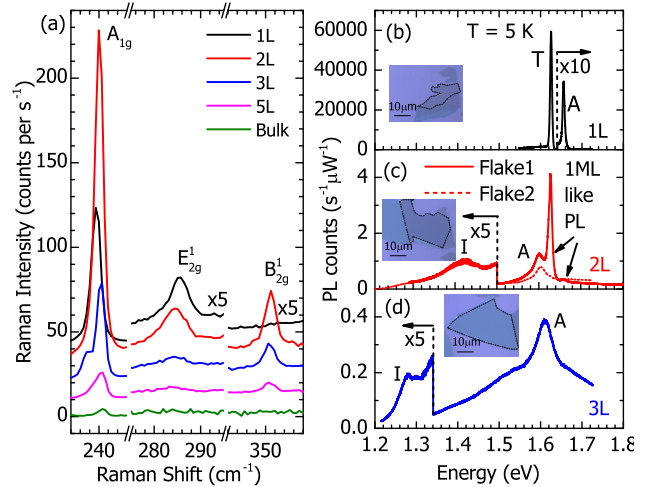}
\caption{(a) Raman spectra obtained for the 1L, 2L, 3L, 5L and 180~nm (bulk) \mose2 flakes. (b) - (d) The low temperature \pl spectra of the 1L, 2L and 3L thick \mose2 flakes with the corresponding optical images displayed as insets. \pl spectra for the 2L flakes appeared to be of two different types, shown by solid (optical image of the flake shown) and dashed lines (image not shown) in (c).}\label{figPL}
\end{figure}

The thicknesses of the mono- to few-layer \mose2 flakes were established using optical contrast, atomic force microscopy (AFM) and Raman spectroscopy techniques. Fig.\ref{figPL}(a) shows the Raman spectra obtained for the flakes under consideration. The Raman lines corresponding to the A$_{1g}$, E$^1_{2g}$ and B$^1_{2g}$ phonon modes were observed in consistence with Ref.~\citenum{oe13pt}. The monolayer thickness was found to be $\sim$0.65~nm using AFM (data not shown). The flakes were then characterized using \pl spectroscopy (see methods). Figure \ref{figPL} (b-d) displays the \pl spectra of 1L to 3L thick flakes. For the 1L flake, two well separated features corresponding to the emission from the n~=~1 ground state A-exciton and the charged exciton (trion, T) are observed. The emission intensity in the case of 2L flake is lower than for the 1L by about 4000 times. A feature corresponding to the indirect gap emission (I) is observed, which is red shifted relative to the direct-gap emission peak A by $\sim$180~meV. Interestingly, in the \pl spectra of some of the 2L flakes (solid line in Fig. \ref{figPL}(b)), we also observed two features which coincide with the A and T emission features of the monolayer, but are around 4 orders of magnitude weaker. These features existed throughout the area of rather large flakes ($\sim$30~$\mu$m wide), therefore the contribution due to the neighbouring 1L flakes can be ruled out.  Our hypothesis is that the process of mechanical exfoliation led to the formation of some pockets within the 2L flakes, which act as two separated monolayers giving rise to a monolayer like emission. The \pl spectrum of the 3L flake was similar to that of the 2L flake where A and I emission features were observed. The I feature was red shifted with respect to A by $\sim$330~meV. We did not observe any measurable emission from the flakes with thicknesses larger than 3Ls.

\section{Excitonic resonances versus layer thickness}
\begin{figure}[!t]
\includegraphics[width=8.5cm]{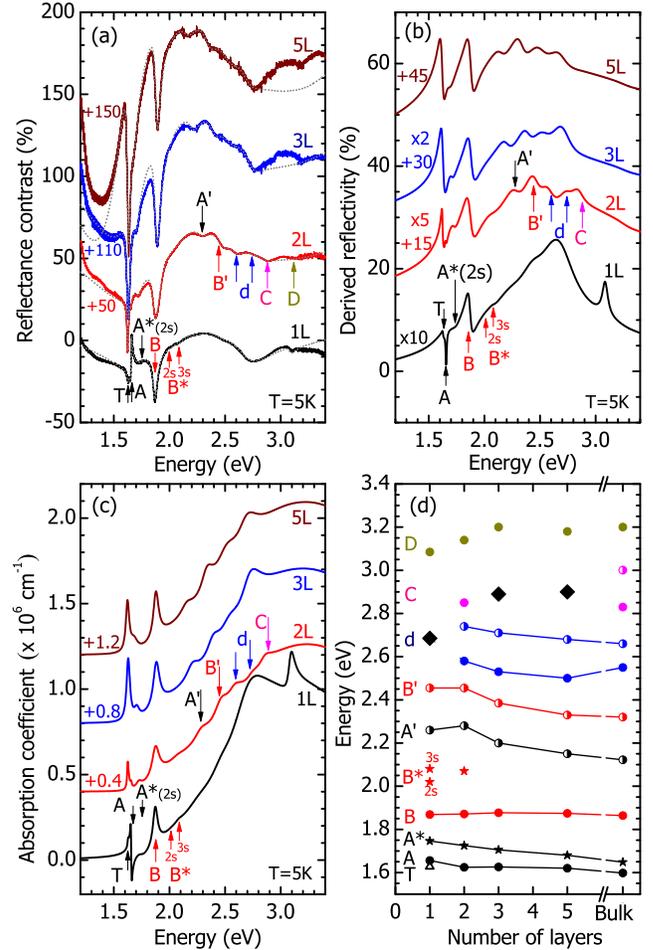}
\caption{(a) Low temperature \rc spectra of the \mose2 flakes (solid lines) and the corresponding modeled curves (dashed grey lines) with thickness ranging from 1L to 5Ls, except a 4L flake. (b) Reflectivity, and (c) the absorption spectra of the flakes derived from the spectral modeling in (a). The features representing A and B excitons, their excited states and the trion (T) are marked for 1L flake, and other excitonic resonances are marked for 2L flake where they are more clearly visible. The spectra for thickness $\ge$2L have been shifted/amplified along \ya for clarity by amounts shown above the curves on the left side. (d) shows the energies of corresponding exciton features as a function of layer thickness. Black diamonds represent the positions of broad features in the spectra, which possibly arise from overlaps between a few closely lying transitions.}\label{figLayerDepRef}
\end{figure}

Figure \ref{figLayerDepRef}(a) shows the reflectance contrast spectra of the \mose2 flakes with layer thicknesses up to 5Ls, except a 4L flake. The spectra are dominated by many excitonic resonance features. To determine the resonance energies, we fitted the spectra using a transfer matrix method which accounted for the interference effects on the exciton lineshapes due to the multilayered structure of the sample as reported earlier.\cite{ns15aa} In our method, the excitonic contribution to the dielectric function was considered to be a modified Lorentzian function\cite{jap13aa} (see Methods section), where a phase factor $\beta$ was used to account for the peculiar lineshape of the A-exciton feature in the case of 1L flake (discussed later), and was set to be equal to 0 for all other features (except A$^*$ for 1L, where $\beta$ was assumed to be equal to that for A-exciton). It is tempting to derive the reflectivity and absorption spectra of the bare flakes (without substrate) from the spectral modeling. These spectra are shown in Fig.~\ref{figLayerDepRef}(b) and (c) respectively where the absorption coefficient $\alpha=4\pi k/\lambda$ is calculated by deriving the imaginary part of the refractive index, $k$, from the modeling. One can clearly observe the features corresponding to the transitions taking place at various points of the Brillouin zone, which have been identified earlier for bulk \mose2 based on low temperature transmission measurements and theoretical calculations.\cite{aip69jaw,jpc72arb} Following these reports, the observed transitions are named as A, B, A$^\prime$, B$^\prime$, d, C and D in the order of increasing energy.
The shoulders A$^*$ and B$^*$ towards the higher energy sides of A and B excitons arise from the contributions due to the excited state exciton (n$\ge$2) transitions. An evolution of the transition energies with the layer thickness is shown in Fig.~\ref{figLayerDepRef}(b). The transition energies for the bulk \mose2 have been obtained from a previous report.\cite{jpc72arb} We notice that contrary to the case of WSe$_2$\cite{ns15aa}, where the A and B exciton resonances display a monotonic red shift when layer thickness is increased, their behavior in \mose2 is different. The A-exciton feature shows an overall red shift ($\sim$58~meV) when the thickness is increased from a monolayer towards bulk, except in going from 2L to 3L where it shows a slight blue shift ($\sim$5~meV). Furthermore, the B-exciton displays an initial blue shift from 1L to 3L ($\sim$8~meV), which is then followed by a red shift ($\sim$13~meV) from 3L to bulk. Similarly, the A$^\prime$ and B$^\prime$ features show an initial blue shift from 1L to 2L thickness, and a red shift for thicker flakes. Although, a detailed theoretical analysis is required to completely understand this phenomenon, it can be qualitatively explained by considering the following two effects: (i) a red shift of the band-to-band absorption edge and (ii) a decreasing excitonic binding energy as the material undergoes a 2D to 3D transition, when the thickness is increased from a monolayer to the bulk limit.\cite{prb84rlg,prl14ac} The excitonic transition appears at an energy lower than the absorption edge by an amount equal to the excitonic binding energy. Therefore, depending upon the magnitudes of the two effects, one can either observe a red or a blue shift of the excitonic transition. In our case of the A-exciton transition, the first effect appears to be much stronger than the second one resulting in a monotonous red shift. However, for B-exciton (A$^\prime$ and B$^\prime$ excitons), the second effect dominates initially up to 3L (2L) thickness resulting in a blue shift, whereas the first one takes over later on which leads to a red shift. For the 1L flake, a broad feature which appears around 2.7~eV (shown by diamond in Fig.~\ref{figLayerDepRef}(b)) is most likely due to unresolved contributions from d- and C- type exciton features. A pair of d-exciton features could be resolved for thicker flakes whereas those related to the C-excitons resulted in a broad feature around 2.8~eV for 3L and 5L flakes (shown by diamonds in Fig.~\ref{figLayerDepRef}(b)). D-exciton features were identified around 3.2~eV for all the flakes but were not accounted for in our modeling. Their positions were assumed to be at their minima and may be subjected to large errors.

Now we discuss the A$^*$ and B$^*$ features in Fig.~\ref{figLayerDepRef} which we associate with the spectral contributions due to the n$\ge$2 excited states of the A and B excitons. The energies of these features agree quite well with those determined recently using 2-photon absorption measurements.\cite{arxiv15gw} Although the A$^*$ feature could be observed for flakes with all thicknesses from 1L to 5L, B$^*$ could only be seen for the 1L and 2L flakes and was too weak to be identified for thicker flakes. Also, only one feature (2s-like) could be identified for the A-exciton for flakes of all thicknesses, whereas for the B-exciton in the 1L flake, both 2s and 3s features could be observed, blue shifted by $\sim$150~meV and $\sim$210~meV respectively with respect to the 1s feature. The exciton series for 1L B-exciton significantly deviates from the 2D Rydberg model for a hydrogen atom in consistence with recent reports on 1L WS$_2$ and WSe$_2$,\cite{prl14kh,prl14ac} which renders it difficult to predict the exciton's binding energy without information about more (n$>$3) excited state features. However, following the ideal 2D Rydberg model for the A-exciton (considering 1s and 2s states) and B-exciton (considering 1s and 3s states), we find that their binding energies should be greater than $~\sim$100~meV and $~\sim$210~meV respectively. In analogy with our earlier findings for WSe$_2$,\cite{ns15aa} the energy difference between the 2s state feature shows an initial blue shift relative to the 1s feature, in going from 1L to 2L flake, for both A and B excitons. This is followed by a red shift in the case of the A-exciton, when the thickness is increased from 2L to 5L. This phenomenon is explained on the basis of two competing effects: (i) the non-local dielectric screening of the exciton excited states for the 1L flake pushes the 2s state closer to 1s state,\cite{prl14kh,prl14ac} and (ii) the material undergoes a 2D to 3D transition while transforming from the 1L to the bulk form. The first effect dominates when the thickness is increased from 1L to 2L whereas the second one takes over for thicker flakes resulting in a red shift.

\section{A-exciton in 1L flake: possible evidence of Fano-resonance}
\begin{figure}[!t]
\includegraphics[width=8.5cm]{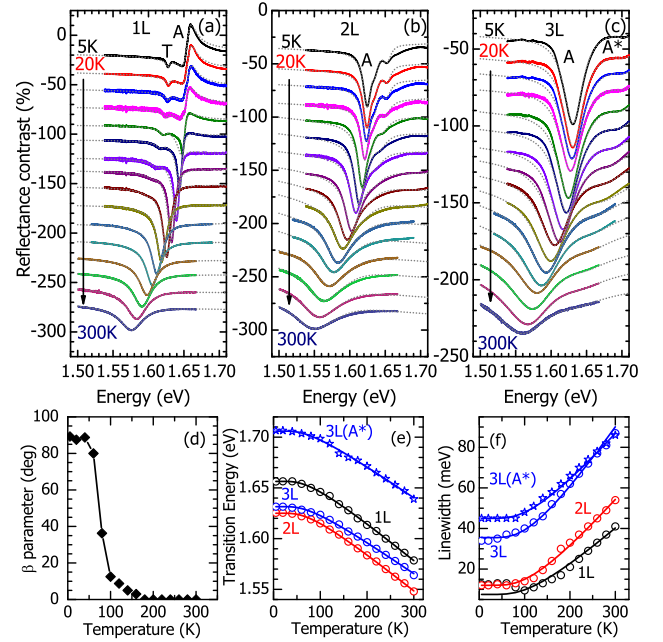}
\caption{(a)-(c) Temperature evolution of the \rc spectra in the spectral region containing A-exciton features for 1L, 2L and 3L \mose2 flakes respectively, along with the modeled curves (dashed grey lines). The spectra for T~$\ge$~20~K have been shifted downwards along \ya for clarity. (d) The $\beta$ parameter for A-exciton in a 1L as derived from the lineshape fits in (a). (e) The transition energy, and (f) the linewidths of the A-exciton features as a function of temperature for 1, 2 and 3~L flakes and A$^*$ feature in 3~L flake along with the fitting (solid lines) based on the models described in the text.}\label{figTDepRef}
\end{figure}

As illustrated in Fig.~\ref{figLayerDepRef}(a to c), our standard analysis of the reflectance spectra (with $\beta=0^\circ$) provides a consistent picture of all observed excitonic transitions but clearly, this does not apply to the case of the A-exciton of the monolayer (with $\beta\sim90^\circ$ at low temperatures, see below). The features related to all (but monolayer A-exciton) imply the absorption resonances in the form of Lorentzian peaks. In contrast, the deduced absorption spectral shape of the 1L A-exciton resonance is unusual, displaying a dispersive-like form. It also shows a negative dip towards the higher energy side of resonance [Fig.~\ref{figLayerDepRef}(c)], but for simplicity, we have not considered any absorption background in our modeling which will result in all positive absorption values. Among known sources of unusual (non-Lorentzian) shapes of excitonic resonances are the effects of significant light-exciton coupling (polaritons) and/or the interaction of the discrete resonance with a continuum of other states (Fano resonance). Polaritonic effects may be relevant in the reflectance spectra of bulk materials \cite{pr58jjh,prb97rs} but can generate only subtle effects in two-dimensional systems \cite{jetp66vma,ssc91lca,spss91eli} unless the latter are embedded into a microcavity,\cite{prl92cw,arxiv15sd} which is not the case of our structures. We thus speculate that the observed unusual shape of the A-exciton in 1L \mose2 is due to Fano-type effect. The excitonic resonances of the Fano shape can indeed be observed in two-dimensional systems. The best known examples are the Fano-type resonances of the excited excitonic states coupled to the continuum of the ground state heavy- and light-hole excitons in GaAs/GaAlAs quantum wells.\cite{prb94dyo,sse96vb,ssc96vb} In search for a continuum of states which possibly couple to the A-exciton in the monolayer (and not in other layers), we note that it is only the monolayer which displays a prominent, below A-exciton absorption peak, related to the charged exciton T. Obviously, there must exist the excited states of the T, including the predicted quasi-continuum of unbound T states which markedly coincide in energy with the undressed neutral exciton resonance.\cite{pssb01ae,pssb01ras} The effective coupling of the A-exciton with the quasi-continuum of excited states of the trion is our hypothesis to account for the unusual shape of the A-exciton in our 1L structure. Temperature dependent studies confirm this conjuncture on a phenomenological ground. As shown in Fig.~\ref{figTDepRef}(a), the ``unusual shape" of the 1L A-exciton transforms into a standard form upon increasing temperature in parallel to the progressive disappearance of the charged exciton resonance. This is accompanied with a gradual reduction of $\beta$ from $\sim$90$^\circ$ to 0$^\circ$. At temperatures above 200~K, when the trion is totally absent in the spectra (and the quasi continuum of its excited states must have disappeared), the A exciton fully recovers its usual shape.

\section{Temperature evolution of absorption and emission spectra}
Fig.~\ref{figTDepRef}(b) and (c) show the temperature evolution of the A-exciton \rc features of the 2L and 3L thick flakes. In the case of a 2L flake at low temperatures (T$<$140~K), the main exciton feature is accompanied with two small features at the higher energy side. Similarly to what has already been observed in the \pl data (Fig.~\ref{figPL}(b)), their energies coincide with the A-exciton and the trion in the 1L flake. For the 3L flake, we also observe the evolution of the A$^*$ feature along with the A-exciton in our spectral window. From modeling of these spectra, we deduce the transition energies and the linewidth broadenings of the excitons as shown in Fig.~\ref{figTDepRef}(e) and (f) respectively. Similar to our results on \wse2,\cite{ns15aa} the temperature activated energy shifts can be equally well fitted using the Varshni\cite{physica67ypv} or O'Donnell \textit{et al.}\cite{apl91kpo} formulae. These relations are given by $E_g(T)=E_0-(\alpha T^2)/(T+\beta)$ and $E_g(T)=E_0-S\langle \hbar\omega \rangle[\coth(\langle \hbar\omega \rangle/2kT)-1]$ respectively. In Varshni's relation, $\alpha$ and $\beta$ are the fitting parameters related to the temperature-dependent dilatation of the lattice and Debye's temperature respectively whereas in O'Donnell's formula, $\langle \hbar\omega \rangle$ and the $S$ are the average phonon energy and coupling parameters respectively. These parameters for the A-exciton in the 1L to 3L flakes and the A$^*$ feature in 3L are given in Table~\ref{tableVarshni} which behave very similarly to those we deduced for \wse2.\cite{ns15aa}
\begin{table}
  \caption{Fitting parameters as obtained from modeling of the resonance energies and lineshape broadenings for the 1L, 2L and 3L ground state A-exciton features, and the A$^*$(2s) feature in 3L flake.}
  \label{tableVarshni}
  \begin{tabular}{p{2.0cm}cccc}
    \hline
    Parameter  & 1L & 2L & 3L & 3L(A$^*$)\\
    \hline
    \multicolumn{5}{c} {Varshni's relation}\\
    \hline
    $E_0$ (eV)  & 1.658 & 1.626 & 1.632 & 1.707\\
    $\alpha (10^{-4} eV/K)$  & 5.67 & 5.56 & 4.80 & 4.77\\
    $\beta$ (K)  & 330 & 330 & 330 & 330\\
    \hline
    \multicolumn{5}{c} {O'Donnell's relation}\\
    \hline
    $E_0$ (eV)  & 1.656 & 1.625 & 1.631 & 1.705\\
    $\langle \hbar\omega \rangle$(meV)  & 19 & 19 & 19 & 19\\
    S  & 2.23 & 2.18 & 1.88 & 1.88\\
    \hline
    \multicolumn{5}{c} {Rudin's relation}\\
    \hline
    $\gamma_0$ (meV)  & 7.5 & 12 & 35.5 & 45\\
    $\gamma^\prime$ (meV)  & 72 & 93 & 120  & 92\\
    \hline
  \end{tabular}
\end{table}
The linewidth broadenings of the A-exciton were fitted using the relation proposed by Rudin et al.\cite{prb90sr}, given as $\gamma(T)=\gamma_0+\sigma T+\gamma^\prime/[1/(e^{\hbar\omega/kT}-1)]$. Here, $\gamma_0$ is the broadening at 0~K, the term linear in T depicts the interaction of excitons with acoustic phonons (disregarded in present work being negligibly small) and the last term arises from the interaction with LO (longitudinal optical) phonons. $\hbar\omega$ is the LO-phonon energy which was taken to be equal to 30~meV.\cite{oe13pt} The fitting parameters for the four features are given in Table~\ref{tableVarshni}. Similar to what has been observed for \wse2,\cite{ns15aa} $\gamma^\prime$ which quantifies the extent of LO-phonon assisted scattering of the ground state exciton with the excited states, increases from 1L to 3L flake. This is because the number of excited states available for scattering increase with increasing flake thickness. However, $\gamma^\prime$ is lower for A$^*$ than for A feature in the 3L flake. It suggests that the excited state feature has lesser number of available bound and continuum excited states for undergoing LO-phonon assisted scattering, as compared to the ground state feature.
\begin{figure}[!t]
\includegraphics[width=7.5cm]{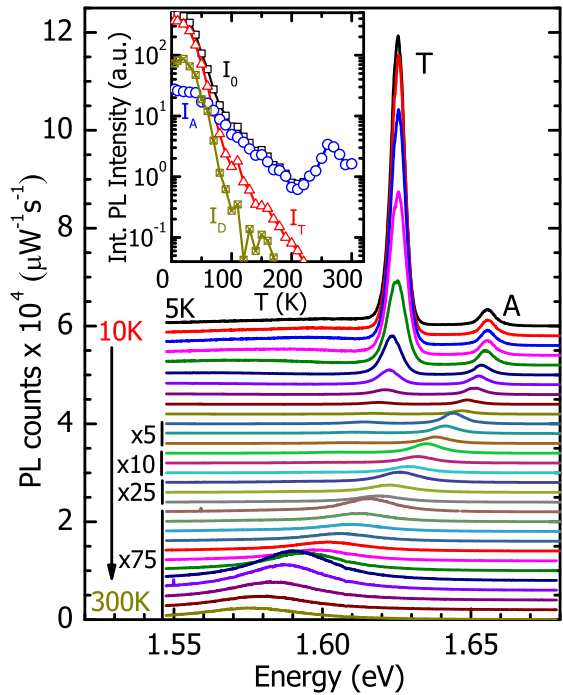}
\caption{Temperature evolution of the \pl spectra of the monolayer \mose2. The spectra for T~$\ge$~10~K have been shifted downwards along \ya for clarity, and those for T$\ge$100~K are amplified with the factors specified. The inset shows the integrated PL intensity for the exciton I$_A$, trion I$_T$, defect related emission I$_D$, and the total emission intensity I$_0$ as a function of temperature.}\label{figTDepPL}
\end{figure}

Finally, we performed \pl as a function of temperature from 5~K till 300~K. The results are shown in Fig.~\ref{figTDepPL}. The inset to the Fig.~\ref{figTDepPL} shows the integrated emission intensity from the trion (I$_T$), the exciton (I$_A$), and the total intensity (I$_0$, area over full spectral range) as a function of temperature. The remaining emission intensity [I$_0$-(I$_A$+I$_T$)] is attributed to the defect related luminescence (I$_D$), which contributes towards the low energy side (below trion) of the spectrum and could result in a slightly asymmetrical trion emission lineshape.\cite{klingshirn} The PL spectrum is initially dominated by trion's contribution at low temperature, which is quickly quenched with increasing temperature and becomes negligible when compared to the exciton's contribution for T$>$200~K. For T$>$70~K, the exciton begins to dominate the spectrum due to quick thermal redistribution of charge carriers in the k-space.\cite{ns15aa} We notice that there is a slight increase in the I$_0$ ($\sim$I$_A$) from 200~K to 250~K (notice log-scale). Although this phenomenon is not well understood at this stage, we note that the change of the PL intensity upon increase of temperature results from the competition between the efficiencies of the radiative and non-radiative recombination channels, the latter being particularly poorly known in the investigated structures. The observed rise of the PL intensity may therefore be a result of distinct temperature dependencies of these two recombination channels. Overall, I$_0$ shows a net drop by a factor of $\sim400$ when temperature was increased from 5~K to 300~K. This is in agreement with another recent report\cite{arxiv15gw2} and is opposite to what we have reported in the case of monolayer \wse2 where I$_0$ showed a gradual increase as a function of temperature.\cite{ns15aa} It can be explained as follows. It has been predicted theoretically that the conduction band (CB) in the monolayers of TMDCs also splits into two bands due to spin-orbit interaction.\cite{prb13kk,2dm15ak} However, the sign of this splitting has been predicted to be opposite for \mose2 and \wse2. Therefore, the states of lower energy among these spin-orbit split CB states have opposite total angular momenta, due to which the lowest energy exciton transitions for \mose2 and \wse2 are expected to be bright and dark respectively.\cite{prb13kk,2dm15ak} Typically, the integrated PL intensity in semiconductors shows a reduction when temperature is increased, due to increasing contribution from the non-radiative recombination processes. If the lowest lying CB state results in the formation of dark excitons in combination with the state at the top of the spin-orbit split valence band (VB), the PL intensity may be quenched at low temperatures where this radiatively inefficient state is preferentially filled with charge carriers. Upon increasing the temperature, the carriers may get transferred to the higher energy CB state upon thermalization, which opens up a radiatively favorable channel for their recombination and may result in an enhancement of emission intensity (which has been observed for \wse2). Therefore, contrary to \wse2, a gradual increase of \pl intensity with increasing temperature is in agreement with the theoretical predictions that the lower energy exciton in \mose2 is bright.

\section{\label{Conclusions}Conclusions}
In conclusion, we studied \mose2 flakes with thicknesses equal to a monolayer, 2Ls, 3Ls and 5Ls using \pl and \rc spectroscopy and as a function of temperature. We observed the evolution of various ground-state resonance features called A, B, A$^\prime$, B$^\prime$, d, C and D as well as the excited-state counterparts of some of them i.e. A$^*$ and B$^*$, as a function of layer thickness. The A-exciton feature in the \rc spectrum of the 1L flake showed Fano-type lineshape at low temperatures. This hypothesis was supported using temperature dependence \rc measurements. In addition, the transition energies and linewidth broadenings of selected excitonic resonances in a 1L to 3L thick flakes were determined as a function of temperature and their behavior was explained using the models routinely used for conventional semiconductors. Finally, our temperature dependent \pl measurements are consistent with the theoretical predictions that the lowest energy exciton in 1L \mose2 is bright.

\section{\label{Methods}Methods}
The monolayer and a few layers \mose2 flakes were obtained on Si/(100~nm)SiO$_2$ substrate by polidimethylosiloxan-based exfoliation\cite{2dm14acg} of bulk crystals purchased from HQ Graphene. The flakes of interest were first identified by visual inspection under an optical microscope.  A transfer matrix based calculation was performed to predict the color of the flake as a function of its thickness.\cite{hecht} The observed color of the flake was then compared with the theoretically predicted one to roughly estimate the number of layers in the flakes. Then the AFM characterization and Raman spectroscopy were performed on the flakes of interest to unambiguously determine their thicknesses. The Raman spectroscopy was performed at room temperature using an argon ion laser (514.5~nm, 0.2~mW focused using a 50x microscope objective), with a 0.5~m focal length monochromator and a 1800 lines/mm grating. For performing \pl measurements, the 632.8~nm radiation from a He-Ne laser was focused on the flake using a 50x long working distance objective. The sample was mounted on the cold finger of a continuous flow liquid helium cryostat, at a temperature of $\sim$5~K. The spot diameter was $\sim$2~$\mu$m and the light power focused on the sample was 1~$\mu$W for the monolayer, 500~$\mu$W for 2L and 2~mW for the 3L flake. The PL emission from the sample was dispersed using a 0.5~m focal length monochromator and detected using a liquid nitrogen cooled Si charge coupled device camera. For performing the \rc measurements, the light from a 100~W tungsten halogen lamp was focused on a pinhole of 75~$\mu$m diameter. The light was then collimated and focused (spot size~$\sim$3$~\mu$m) on the sample. The reflected light was detected using a setup similar to the one used for performing \pl. If $\mathcal{R}(\lambda)$ and $\mathcal{R}_0(\lambda)$ are the wavelength dependent reflectance spectra of the \mose2 flake and of the Si/SiO$_2$ substrate respectively, then the percentage reflectance contrast spectrum $\mathcal{C}(\lambda)$ was determined using $\mathcal{C}(\lambda)=[\mathcal{R}(\lambda)-\mathcal{R}_0(\lambda)]/[\mathcal{R}(\lambda)+\mathcal{R}_0(\lambda)]$. The measurements were performed at temperatures ranging from 5~K to 300~K. For the lineshape analysis of the RC spectra, we followed a method similar to that described in Ref.~\citenum{jap13aa}. In this method, we considered the excitonic contribution to the dielectric response function to be given by a modified Lorentz oscillator like model as
\begin{equation}
\varepsilon (E)=(n_b + ik_b)^2 + \sum_p{\frac{A_p e^{\i \beta}}{E_{p}^2-E^2-i\gamma_p E}},\label{eqnLor}
\end{equation}
where $n_b + ik_b$ represents the background complex refractive index of \mose2 in the absence of excitons and was assumed to be equal to that of the bulk material\cite{jpc79arb} in the simulations. The index $p$ stands for the type of exciton characterized by a resonance energy $E_{p}$, an amplitude $A_p$, a phase factor $\beta$ (to account for the Fano-type lineshape) and a phenomenological broadening parameter $\gamma_p$ [equal to full width at half maximum (FWHM) of the Lorentzian function]. The RC was then calculated using the transfer matrix formalism. It may be mentioned that despite storing the samples in a vacuum desiccator, we noticed an overall drop in the PL intensity nearly by a factor of 3 when the PL measurements were repeated on the monolayer flake after 40 days which indicates the possible effects of sample-aging.

\section{Author contributions}
A.A. performed the optical measurements, analyzed the data and did most of the manuscript writing with significant inputs from K. N. and M. P. The \mose2 layers were prepared and identified by K. N. and M. M. helped determining the layer thickness using Raman spectroscopy measurements. M. K. participated in optical spectroscopy measurements. M. P. supervised the project and provided critical insights into data interpretation and manuscript content.

\begin{acknowledgement}
We thank Ivan Breslavetz for technical assistance and acknowledge support from the EC Graphene Flagship project (No. 604391), the European Research Council (MOMB project No. 320590), the NCN 2013/10/M/ST3/00791 grant and the Nanofab facility of the Institut N\'{e}el, CNRS-UGA. A. A. thanks Shreya Agrawal for support during experimental setup.
\end{acknowledgement}

\end{document}